\documentclass[12pt]{article}
\usepackage[utf8]{inputenc}
\usepackage{amsmath}
\usepackage{amssymb}
\usepackage[a4paper]{geometry}
\usepackage[bookmarks=false,
 breaklinks=false,pdfborder={0 0 1},backref=section,colorlinks=false]
 {hyperref}
\hypersetup{
 hidelinks}

\makeatletter

\usepackage{cite}
\usepackage{authblk}
\usepackage{slashed}

\allowdisplaybreaks

\newcommand{\Ep}{E_{\mathrm{p}}}
\newcommand{\ii}{\mathrm{i}}
\newcommand{\dd}{\mathrm{d}}
\usepackage{orcidlink}

\title{The Deformed Dirac Oscillator in Linear-Fractional Doubly Special Relativity}

\author[1,2,3]{Nosratollah Jafari\,\orcidlink{0000-0002-0285-6107}\thanks{Email: \texttt{nosrat.jafari@fai.kz}}}

\author[4]{Abdelmalek Boumali\,\orcidlink{0000-0003-2552-0427}\thanks{Email: \texttt{boumali.abdelmalek@gmail.com}}}

\affil[1]{Fesenkov Astrophysical Institute, 050020, Almaty, Kazakhstan}
\affil[2]{Al-Farabi Kazakh National University, Al-Farabi av. 71, 050040 Almaty, Kazakhstan}
\affil[3]{Center for Theoretical Physics, Khazar University, 41 Mehseti Street, Baku, AZ1096, Azerbaijan}
\affil[4]{Laboratoire de Physique Appliqu\'ee et Th\'eorique (LPAT), Universit\'e de T\'ebessa, 12000 T\'ebessa, Algeria}

\date{\today}

\makeatother

\begin{document}
\maketitle


\begin{abstract}
We study the $(1+1)$-dimensional Dirac oscillator within a class
of doubly special relativity (DSR) models generated by linear-fractional
(projective) transformations on momentum space that preserve both
the invariant speed of light and a high-energy observer-independent
scale $\Ep$. Starting from the associated deformed Casimir invariants,
we construct the coordinate-space Dirac equations for three inequivalent
choices of the deformation vector (time-like, space-like, and light-like).
For the time-like and light-like realizations the deformation induces
momentum-dependent effective mass operators, which makes the coordinate-space
formulation sensitive to operator ordering. To retain locality and
obtain solvable second-order equations we adopt a reverted-product
ordering prescription. Closed-form relativistic energy spectra and
eigenfunctions are obtained in all three geometries, and the standard
Dirac-oscillator results are recovered smoothly in the limit $\Ep\to\infty$.
Finally, we derive the nonrelativistic expansion of the positive-energy
branch and show that the deformation geometry controls the leading
rest-energy shift and the renormalization of the oscillator level
spacing. 
\end{abstract}


\section{Introduction}

Doubly special relativity (DSR) extends special-relativistic kinematics
by postulating, in addition to the invariant speed of light, a second
observer-independent scale---typically an energy scale $\Ep$ associated
with Planckian physics. This hypothesis leads to deformed dispersion
relations and nonlinear realizations of Lorentz transformations in
momentum space \cite{AmelinoCamelia2002,MagueijoSmolin2002,JudesVisser2003}.
In many developments, DSR-inspired kinematics is naturally interpreted
in geometric terms, most notably through curved momentum space and
related phase-space formulations of the relativity principle. A mathematically
transparent realization of several DSR proposals is provided by projective
(linear-fractional) actions on momentum space, which organize seemingly
different deformations into a common geometric structure \cite{JafariShariati2011};
contemporary work further clarifies the relation between DSR, $\kappa$-Poincaré
structures, and momentum-space curvature \cite{Jafari2024Curvature}.

The Dirac oscillator (DO) is a paradigmatic exactly solvable relativistic
model obtained by a non-minimal, linear-in-coordinate modification
of the Dirac momentum operator. Early appearances of oscillator-like
Dirac dynamics can be traced to the work of Itô, Mori and Carrière
\cite{Ito1967}, while the modern covariant formulation known as the
Dirac--Moshinsky oscillator was popularized by Moshinsky and Szczepaniak
\cite{Moshinsky1989} (see also \cite{MorenoZentella1989}). The DO
exhibits a rich algebraic structure, including hidden supersymmetry
\cite{Benitez1990}, and it provides a bridge between relativistic
quantum mechanics and quantum-optics models: in $(2+1)$ dimensions
it admits an exact mapping onto the Jaynes--Cummings interaction,
enabling ion-trap simulation proposals \cite{Bermudez2007}. The model
and its extensions have been applied in a wide range of contexts,
including relativistic confinement, Dirac materials, and thermodynamic/statistical
investigations; see, e.g., the overview \cite{Sadurni2011} and representative
studies \cite{BoumaliHassanabadi2013,Boumali2015PhysScr,BoumaliHassanabadi2015ZNA,BoumaliChetouaniHassanabadi2016APPB,BoumaliHassanabadi2017AHEP}.

Motivated by the use of the DO as a probe of modified relativistic
kinematics, recent work has examined oscillator dynamics in DSR and
related deformed frameworks. Comparative analyses of the DO in the
Magueijo--Smolin and Amelino--Camelia DSR models, including controlled
nonrelativistic expansions, were reported in \cite{JafariBoumali2025},
while thermodynamic properties of DSR-modified oscillators were studied
in \cite{BoumaliEtAl2025EPJC,BoumaliJafari2026CTP}. Related DSR wave-equation
problems with confining interactions have also been investigated;
see, for example, \cite{SeffaiMeradHamil2022}.

\paragraph{Objective and outline.}

The objective of this paper is to provide an analytic and geometry-sensitive
treatment of the one-dimensional Dirac oscillator in a projective
(linear-fractional) DSR setting. Specifically, we: (i) formulate the
deformed mass shell induced by a general projective momentum map and
derive the associated coordinate-space Dirac equations; (ii) implement
the Dirac-oscillator coupling for time-like, space-like, and light-like
deformation geometries; (iii) resolve the operator-ordering ambiguity
that arises in the presence of momentum-dependent effective mass operators;
and (iv) obtain exact relativistic spectra and eigenfunctions, as
well as the nonrelativistic limit of the positive-energy branch. Section~2
develops the formalism and presents the exact solutions, and Section~3
summarizes the main results and their physical implications.

\section{Linear-fractional DSR Dirac oscillator}

\subsection{Projective momentum map and modified mass shell}

We consider a projective map $f$ acting on momentum space, 
\begin{equation}
f:\quad p^{\mu}\longmapsto\pi^{\mu}=\frac{p^{\mu}}{1+\Ep^{-1}\,\eta_{\alpha\beta}a^{\alpha}p^{\beta}},
\end{equation}
where $\eta_{\mu\nu}=\mathrm{diag}(-1,1)$ in $(1+1)$ dimensions
and $a^{\mu}$ is a constant deformation vector. The auxiliary variables
$\pi^{\mu}$ transform linearly under the Lorentz group, while the
physical momenta $p^{\mu}$ transform under the induced nonlinear
action $f^{-1}\Lambda f$ \cite{JafariShariati2011}. The corresponding
linear-fractional Lorentz transformation reads 
\begin{equation}
p'{}^{\mu}=\frac{\Lambda_{\ \nu}^{\mu}p^{\nu}}{1+\Ep^{-1}a_{\alpha}(\delta_{\ \beta}^{\alpha}-\Lambda_{\ \beta}^{\alpha})p^{\beta}}.\label{eq:lf-general}
\end{equation}
The associated invariant mass shell (deformed Casimir) can be written
as 
\begin{equation}
\frac{\eta_{\mu\nu}p^{\mu}p^{\nu}}{\left(1+\Ep^{-1}\,\eta_{\alpha\beta}a^{\alpha}p^{\beta}\right)^{2}}=m^{2},\label{eq:casimir-general}
\end{equation}
which reduces to the standard relativistic dispersion relation as
$\Ep\to\infty$.

\subsection{Geometric deformations and linearized Dirac equations}

To construct first-order wave equations we linearize the deformed
dispersion relation by associating the numerator with the Dirac operator
and the denominator with an effective, deformation-dependent mass
factor. We employ the correspondence $p_{\mu}\to-\ii\partial_{\mu}$
and use the representation $\gamma^{0}=\sigma_{3}$, $\gamma^{1}=\ii\sigma_{2}$.

\subsubsection*{(i) Time-like deformation (Magueijo--Smolin type)}

The deformed mass shell is taken as 
\begin{equation}
\frac{E^{2}-p^{2}}{(1-E/\Ep)^{2}}=m^{2},\label{eq:casimir-timelike}
\end{equation}
which linearizes to $\gamma^{\mu}p_{\mu}=m\left(1-E/\Ep\right)$.
In coordinate space this yields 
\begin{equation}
\left[\,\ii\gamma^{\mu}\partial_{\mu}-m\left(1-\frac{\ii}{\Ep}\frac{\partial}{\partial t}\right)\right]\Psi=0.\label{eq:dirac-timelike}
\end{equation}

\subsubsection*{(ii) Space-like deformation}

We take 
\begin{equation}
\frac{E^{2}-p^{2}}{(1-p/\Ep)^{2}}=m^{2},\label{eq:casimir-spacelike}
\end{equation}
so that $\gamma^{\mu}p_{\mu}=m\left(1-p/\Ep\right)$. Using $p\to-\ii\partial_{x}$
gives 
\begin{equation}
\left[\,\ii\gamma^{\mu}\partial_{\mu}-m\left(1+\frac{\ii}{\Ep}\frac{\partial}{\partial x}\right)\right]\Psi=0.\label{eq:dirac-spacelike}
\end{equation}

\subsubsection*{(iii) Light-like (null) deformation}

For the null case we use 
\begin{equation}
\frac{E^{2}-p^{2}}{\left(1-(E+p)/\Ep\right)^{2}}=m^{2},\label{eq:casimir-lightlike}
\end{equation}
which linearizes to $\gamma^{\mu}p_{\mu}=m\left[1-(E+p)/\Ep\right]$
and hence 
\begin{equation}
\left[\,\ii\gamma^{\mu}\partial_{\mu}-m\left(1-\frac{\ii}{\Ep}\frac{\partial}{\partial t}+\frac{\ii}{\Ep}\frac{\partial}{\partial x}\right)\right]\Psi=0.\label{eq:dirac-lightlike}
\end{equation}

\subsection{Dirac-oscillator coupling and ordering}

The one-dimensional Dirac oscillator is introduced through the non-minimal
substitution \cite{Moshinsky1989} 
\begin{equation}
p\longrightarrow\Pi=p-\ii m\omega\,\beta x,\qquad\beta\equiv\gamma^{0},\qquad p=-\ii\frac{\dd}{\dd x}.\label{eq:do-subst}
\end{equation}
For stationary states $\Psi(x,t)=e^{-\ii Et}\psi(x)$ and $\psi=(\psi_{1},\psi_{2})^{T}$,
the time-like geometry yields an energy-dependent but $x$-independent
effective mass $M_{t}(E)=m(1-E/\Ep)$. In contrast, the space-like
and light-like realizations generate momentum-dependent effective
mass operators 
\begin{equation}
M_{s}(p)=m\left(1-\frac{p}{\Ep}\right),\qquad M_{\ell}(E,p)=m\left(1-\frac{E}{\Ep}-\frac{p}{\Ep}\right),
\end{equation}
which do not commute with $x$ once $p=-\ii\partial_{x}$.

To maintain a local second-order equation after decoupling the Dirac
system in coordinate space, we adopt an ordered (reverted-product)
prescription, schematically 
\begin{equation}
(E+M(p))\,(p\pm\ii m\omega x)\ \longrightarrow\ (p\pm\ii m\omega x)\,(E+M(p)),\label{eq:ordered}
\end{equation}
consistent with the ordering used in related DSR oscillator problems
\cite{JafariBoumali2025,SeffaiMeradHamil2022}.

\subsection{Exact solutions and nonrelativistic limits (time-like, space-like,
light-like)}

\label{subsec:exact-solutions}

We now solve the deformed Dirac-oscillator problem for the three deformation
geometries. We use the stationary ansatz $\Psi(x,t)=e^{-\ii Et}\psi(x)$
with $\psi(x)=(\psi_{1}(x),\psi_{2}(x))^{T}$, the representation
$\gamma^{0}=\sigma_{3}$, $\gamma^{1}=\ii\sigma_{2}$, and the Dirac-oscillator
substitution $p\to\Pi=p-\ii m\omega\,\beta x$ with $\beta=\gamma^{0}$.
In all cases we write $p=-\ii\,\frac{\dd}{\dd x}$.

For a given deformation geometry, the stationary Dirac equation can
be written in the form 
\begin{equation}
(E-\mathcal{M})\psi_{1}=(p+\ii m\omega x)\psi_{2},\qquad(E+\mathcal{M})\psi_{2}=(p-\ii m\omega x)\psi_{1},\label{eq:generic-coupled}
\end{equation}
where $\mathcal{M}$ is the effective mass (a c-number in the time-like
case and an operator in the space-like/light-like cases). To obtain
a local second-order equation we decouple \eqref{eq:generic-coupled}.
In the operator-valued cases we apply the ordered (reverted-product)
rule \eqref{eq:ordered} when moving $(E+\mathcal{M})$ past $(p\pm\ii m\omega x)$.

A key identity used repeatedly is 
\begin{align}
(p+\ii m\omega x)(p-\ii m\omega x) & =p^{2}+m^{2}\omega^{2}x^{2}+\ii m\omega[x,p]\nonumber \\
 & =p^{2}+m^{2}\omega^{2}x^{2}-m\omega,\label{eq:key-identity}
\end{align}
since $[x,p]=\ii$.


\subsubsection*{(i) Time-like deformation (Magueijo--Smolin): exact spectrum and
NR limit}

For the time-like deformation, the stationary equation implies the
energy-dependent (but $x$-independent) effective mass 
\begin{equation}
\mathcal{M}\equiv M_{t}(E)=m\left(1-\frac{E}{\Ep}\right).
\end{equation}
Since $M_{t}$ is a c-number for each eigenvalue $E$, ordering issues
do not arise. Starting from \eqref{eq:generic-coupled}, multiply
the first equation by $(E+M_{t})$ and substitute the second equation:
\begin{align}
(E+M_{t})(E-M_{t})\psi_{1} & =(E+M_{t})(p+\ii m\omega x)\psi_{2}\nonumber \\
 & =(p+\ii m\omega x)(E+M_{t})\psi_{2}\nonumber \\
 & =(p+\ii m\omega x)(p-\ii m\omega x)\psi_{1}.
\end{align}
Thus, 
\begin{equation}
\left[(p+\ii m\omega x)(p-\ii m\omega x)-\left(E^{2}-M_{t}^{2}\right)\right]\psi_{1}=0.
\end{equation}
Using \eqref{eq:key-identity} we obtain the harmonic-oscillator form
\begin{equation}
\left[p^{2}+m^{2}\omega^{2}x^{2}-m\omega\right]\psi_{1}(x)=\left(E^{2}-M_{t}^{2}\right)\psi_{1}(x).\label{eq:timelike-secondorder-expanded}
\end{equation}
The standard oscillator eigenvalue relation is 
\begin{equation}
\left[p^{2}+m^{2}\omega^{2}x^{2}\right]\phi_{n}(x)=m\omega(2n+1)\phi_{n}(x),\qquad n=0,1,2,\dots
\end{equation}
so \eqref{eq:timelike-secondorder-expanded} implies 
\begin{equation}
E^{2}-m^{2}\left(1-\frac{E}{\Ep}\right)^{2}=2m\omega\,n.\label{eq:quant-timelike}
\end{equation}
Solving \eqref{eq:quant-timelike} yields the exact spectrum 
\begin{equation}
E_{n}^{(\pm)}=\frac{-m^{2}/\Ep\pm\sqrt{\,m^{2}+2m\omega n\left(1-\frac{m^{2}}{\Ep^{2}}\right)\,}}{1-\frac{m^{2}}{\Ep^{2}}}.\label{eq:spectrum-timelike}
\end{equation}

\paragraph{Eigenfunctions.}

Up to normalization, 
\begin{equation}
\psi_{1,n}(x)\propto\exp\!\left(-\frac{m\omega}{2}x^{2}\right)\,H_{n}\!\left(\sqrt{m\omega}\,x\right),
\end{equation}
and the lower component follows from \eqref{eq:generic-coupled}:
\begin{equation}
\psi_{2,n}(x)=\frac{1}{E_{n}^{(+)}+M_{t}(E_{n}^{(+)})}\,(p-\ii m\omega x)\,\psi_{1,n}(x)\quad\text{(positive branch),}
\end{equation}
with an analogous expression for the negative branch.

\paragraph{Nonrelativistic limit (positive-energy branch).}

Write $E_{n}^{(+)}=m+\varepsilon_{n}$ with $\varepsilon_{n}\ll m$
and assume $m/\Ep\ll1$. Expanding \eqref{eq:spectrum-timelike} gives
\begin{equation}
E_{n}^{(+)}=m-\frac{m^{2}}{\Ep}+\frac{m^{3}}{\Ep^{2}}+\omega n-\frac{\omega^{2}n^{2}}{2m}+\frac{m\,\omega^{2}n^{2}}{2\Ep^{2}}+\mathcal{O}\!\left(\frac{\omega^{3}}{m^{2}},\frac{1}{\Ep^{3}}\right).\label{eq:NR-timelike}
\end{equation}
Hence, in the time-like geometry the leading Planck-suppressed effect
is a rest-energy shift $\sim-m^{2}/\Ep$, while the leading oscillator
spacing remains $\varepsilon_{n}\simeq\omega n$.


\subsubsection*{(ii) Space-like deformation: exact spectrum, phase removal, and NR
limit}

For the space-like deformation the stationary equation leads to a
momentum-dependent mass operator, 
\begin{equation}
\mathcal{M}\equiv M_{s}(p)=m\left(1-\frac{p}{\Ep}\right).
\end{equation}
Starting again from \eqref{eq:generic-coupled}, decoupling with the
ordering rule \eqref{eq:ordered} yields 
\begin{equation}
\left[(p+\ii m\omega x)(p-\ii m\omega x)-E^{2}+M_{s}^{2}\right]\psi_{1}=0.\label{eq:spacelike-decouple-step}
\end{equation}
Using \eqref{eq:key-identity} and 
\begin{equation}
M_{s}^{2}=m^{2}\left(1-\frac{p}{\Ep}\right)^{2}=m^{2}\left(1-\frac{2p}{\Ep}+\frac{p^{2}}{\Ep^{2}}\right),
\end{equation}
we obtain 
\begin{equation}
\left[\left(1+\frac{m^{2}}{\Ep^{2}}\right)p^{2}-\frac{2m^{2}}{\Ep}\,p+m^{2}\omega^{2}x^{2}+\left(m^{2}-m\omega-E^{2}\right)\right]\psi_{1}(x)=0.\label{eq:spacelike-pre-gauge-extended}
\end{equation}
The linear term in $p$ corresponds to a first-derivative term in
coordinate space. It can be removed by a constant phase (gauge) transformation
\begin{equation}
\psi_{1}(x)=e^{\ii\beta_{s}x}\,\chi_{1}(x),\qquad\beta_{s}=\frac{m^{2}}{\Ep\left(1+\frac{m^{2}}{\Ep^{2}}\right)}.\label{eq:psi1-gauge-spacelike}
\end{equation}
Indeed, $p\psi_{1}=e^{\ii\beta_{s}x}(p+\beta_{s})\chi_{1}$ and $p^{2}\psi_{1}=e^{\ii\beta_{s}x}(p+\beta_{s})^{2}\chi_{1}$;
with the choice \eqref{eq:psi1-gauge-spacelike}, the coefficient
of the linear-$p$ term vanishes and \eqref{eq:spacelike-pre-gauge-extended}
reduces to 
\begin{equation}
\left[\left(1+\frac{m^{2}}{\Ep^{2}}\right)p^{2}+m^{2}\omega^{2}x^{2}\right]\chi_{1}=\Lambda_{s}\,\chi_{1},\label{eq:spacelike-reduced-extended}
\end{equation}
with 
\begin{equation}
\Lambda_{s}=E^{2}-m^{2}+m\omega+\frac{m^{4}}{\Ep^{2}+m^{2}}.\label{eq:Lambda-s-extended}
\end{equation}
Dividing \eqref{eq:spacelike-reduced-extended} by $\left(1+\frac{m^{2}}{\Ep^{2}}\right)$,
we identify a harmonic oscillator with effective frequency 
\begin{equation}
\omega_{\mathrm{eff}}=\frac{\omega}{\sqrt{1+\frac{m^{2}}{\Ep^{2}}}},
\end{equation}
and hence 
\begin{equation}
\frac{\Lambda_{s}}{1+\frac{m^{2}}{\Ep^{2}}}=m\omega_{\mathrm{eff}}(2n+1)=\frac{m\omega}{\sqrt{1+\frac{m^{2}}{\Ep^{2}}}}(2n+1).
\end{equation}
Therefore, 
\begin{equation}
\Lambda_{s}=m\omega\sqrt{1+\frac{m^{2}}{\Ep^{2}}}\,(2n+1),
\end{equation}
and inserting \eqref{eq:Lambda-s-extended} gives the exact spectrum
\begin{equation}
E_{n}^{(\pm)}=\pm\sqrt{m^{2}-m\omega+m\omega\sqrt{1+\frac{m^{2}}{\Ep^{2}}}\,(2n+1)-\frac{m^{4}}{\Ep^{2}+m^{2}}}.\label{eq:spectrum-spacelike}
\end{equation}

\paragraph{Eigenfunctions.}

Up to normalization, 
\begin{equation}
\chi_{1,n}(x)\propto\exp\!\left(-\frac{m\omega_{\mathrm{eff}}}{2}x^{2}\right)\,H_{n}\!\left(\sqrt{m\omega_{\mathrm{eff}}}\,x\right),\qquad\psi_{1,n}(x)=e^{\ii\beta_{s}x}\chi_{1,n}(x),
\end{equation}
and $\psi_{2,n}$ follows from \eqref{eq:generic-coupled}.

\paragraph{Nonrelativistic limit (positive-energy branch).}

Expanding \eqref{eq:spectrum-spacelike} for $m/\Ep\ll1$ and $\omega n\ll m$
gives 
\begin{equation}
E_{n}^{(+)}=m-\frac{m^{3}}{2\Ep^{2}}+\omega n\left(1+\frac{m^{2}}{\Ep^{2}}\right)+\frac{m^{2}\omega}{4\Ep^{2}}-\frac{\omega^{2}n^{2}}{2m}-\frac{m^{2}}{\Ep^{2}}\,\frac{\omega^{2}\,n(5n+1)}{4m}+\mathcal{O}\!\left(\frac{\omega^{3}}{m^{2}},\frac{1}{\Ep^{3}}\right).\label{eq:NR-spacelike}
\end{equation}
Thus, in the space-like geometry the leading deformation in the excitation
energies is a spacing renormalization $\omega\to\omega\left(1+\frac{m^{2}}{\Ep^{2}}\right)$,
while there is no $\mathcal{O}(1/\Ep)$ rest-energy shift.


\subsubsection*{(iii) Light-like (null) deformation: quadratic quantization, exact
spectrum, and NR limit}

For the light-like deformation, the stationary effective mass depends
on both $E$ and $p$: 
\begin{equation}
\mathcal{M}\equiv M_{\ell}(E,p)=m\left(1-\frac{E}{\Ep}-\frac{p}{\Ep}\right)=m\left(\alpha-\frac{p}{\Ep}\right),\qquad\alpha:=1-\frac{E}{\Ep}.
\end{equation}
Decoupling \eqref{eq:generic-coupled} with the ordering rule \eqref{eq:ordered}
gives 
\begin{equation}
\left[(p+\ii m\omega x)(p-\ii m\omega x)-E^{2}+M_{\ell}^{2}\right]\psi_{1}=0.\label{eq:lightlike-decouple-step}
\end{equation}
Using \eqref{eq:key-identity} and 
\begin{equation}
M_{\ell}^{2}=m^{2}\left(\alpha-\frac{p}{\Ep}\right)^{2}=m^{2}\left(\alpha^{2}-\frac{2\alpha}{\Ep}p+\frac{p^{2}}{\Ep^{2}}\right),
\end{equation}
we obtain 
\begin{equation}
\left[\left(1+\frac{m^{2}}{\Ep^{2}}\right)p^{2}-\frac{2m^{2}\alpha}{\Ep}\,p+m^{2}\omega^{2}x^{2}+\left(m^{2}\alpha^{2}-m\omega-E^{2}\right)\right]\psi_{1}(x)=0.\label{eq:lightlike-pre-gauge-extended}
\end{equation}
Remove the linear-$p$ term by 
\begin{equation}
\psi_{1}(x)=e^{\ii\beta_{\ell}x}\,\chi_{1}(x),\qquad\beta_{\ell}=\frac{m^{2}\alpha}{\Ep\left(1+\frac{m^{2}}{\Ep^{2}}\right)}=\frac{m^{2}\left(1-\frac{E}{\Ep}\right)}{\Ep\left(1+\frac{m^{2}}{\Ep^{2}}\right)}.\label{eq:psi1-gauge-lightlike}
\end{equation}
This reduces \eqref{eq:lightlike-pre-gauge-extended} to 
\begin{equation}
\left[\left(1+\frac{m^{2}}{\Ep^{2}}\right)p^{2}+m^{2}\omega^{2}x^{2}\right]\chi_{1}=\Lambda_{\ell}\,\chi_{1},\label{eq:lightlike-reduced-extended}
\end{equation}
where 
\begin{equation}
\Lambda_{\ell}=E^{2}+m\omega-m^{2}\left(1-\frac{E}{\Ep}\right)^{2}+\frac{m^{4}\left(1-\frac{E}{\Ep}\right)^{2}}{\Ep^{2}+m^{2}}.\label{eq:Lambda-l-extended}
\end{equation}
Since the operator on the left-hand side of \eqref{eq:lightlike-reduced-extended}
is the same oscillator operator as in the space-like case, its eigenvalues
are $m\omega\sqrt{1+\frac{m^{2}}{\Ep^{2}}}(2n+1)$, yielding the light-like
quantization condition 
\begin{equation}
E^{2}+m\omega-m^{2}\left(1-\frac{E}{\Ep}\right)^{2}+\frac{m^{4}\left(1-\frac{E}{\Ep}\right)^{2}}{\Ep^{2}+m^{2}}=m\omega\sqrt{1+\frac{m^{2}}{\Ep^{2}}}\,(2n+1).\label{eq:lightlike-quant-extended}
\end{equation}
Equation \eqref{eq:lightlike-quant-extended} is quadratic in $E$
and can be solved in closed form, giving the exact spectrum 
\begin{equation}
E_{n}^{(\pm)}=-\frac{m^{2}}{\Ep}\pm\frac{\sqrt{\Ep^{2}+m^{2}}}{\Ep}\,\sqrt{\,m^{2}-m\omega+m\omega\sqrt{1+\frac{m^{2}}{\Ep^{2}}}\,(2n+1)\,}.\label{eq:spectrum-lightlike}
\end{equation}

\paragraph{Eigenfunctions.}

The upper component is again an oscillator Hermite function multiplied
by the phase $e^{\ii\beta_{\ell}x}$: 
\begin{equation}
\chi_{1,n}(x)\propto\exp\!\left(-\frac{m\omega_{\mathrm{eff}}}{2}x^{2}\right)\,H_{n}\!\left(\sqrt{m\omega_{\mathrm{eff}}}\,x\right),\qquad\psi_{1,n}(x)=e^{\ii\beta_{\ell}x}\chi_{1,n}(x),
\end{equation}
with $\omega_{\mathrm{eff}}=\omega/\sqrt{1+m^{2}/\Ep^{2}}$ as before,
but now $\beta_{\ell}$ depends on the eigenvalue $E$ through \eqref{eq:psi1-gauge-lightlike}.
The lower component follows from \eqref{eq:generic-coupled}.

\paragraph{Nonrelativistic limit (positive-energy branch).}

Expanding \eqref{eq:spectrum-lightlike} for $m/\Ep\ll1$ and $\omega n\ll m$
gives 
\begin{equation}
E_{n}^{(+)}=m-\frac{m^{2}}{\Ep}+\frac{m^{3}}{2\Ep^{2}}+\omega n\left(1+\frac{m^{2}}{\Ep^{2}}\right)+\frac{m^{2}\omega}{4\Ep^{2}}-\frac{\omega^{2}n^{2}}{2m}-\frac{m^{2}}{\Ep^{2}}\,\frac{\omega^{2}\,n(3n+1)}{4m}+\mathcal{O}\!\left(\frac{\omega^{3}}{m^{2}},\frac{1}{\Ep^{3}}\right).\label{eq:NR-lightlike}
\end{equation}
Thus, the light-like geometry combines the MS-type $\mathcal{O}(m^{2}/\Ep)$
rest-energy shift with the space-like-type $\mathcal{O}(m^{2}/\Ep^{2})$
renormalization of the leading level spacing.

\paragraph{Geometric comparison (within this solvable model).}

At the level of the nonrelativistic excitation energies, the dominant
effects are: (i) time-like deformation mainly shifts the rest energy
at order $1/\Ep$ and leaves the leading spacing $\sim\omega n$ unchanged;
(ii) space-like deformation produces a spacing renormalization at
order $1/\Ep^{2}$ without a linear $1/\Ep$ shift; (iii) light-like
deformation exhibits both features.


\section{Conclusion}

We investigated the one-dimensional Dirac oscillator in a class of
projective (linear-fractional) DSR models characterized by an invariant
energy scale $\Ep$. Starting from the corresponding deformed Casimir
invariants, we derived the coordinate-space Dirac equations in $(1+1)$
dimensions for three inequivalent deformation geometries (time-like,
space-like, and light-like) and implemented the Dirac-oscillator coupling.

A central technical issue is that, for the time-like and light-like
realizations, the deformation promotes the mass term to a momentum-dependent
operator, which introduces an operator-ordering ambiguity once one
passes to coordinate space. Adopting a reverted-product ordering prescription,
we obtained local second-order equations and derived closed-form spectra
and eigenfunctions in all three geometries. In each case the ordinary
Dirac oscillator is recovered smoothly in the undeformed limit $\Ep\to\infty$.

The nonrelativistic expansion of the positive-energy branch highlights
how the deformation geometry controls the leading corrections. In
the time-like case, the dominant effect is a linear $\mathcal{O}(m^{2}/\Ep)$
shift of the rest energy, while the leading oscillator spacing remains
unchanged at this order. In contrast, the space-like deformation produces
no linear $1/\Ep$ rest-energy shift but renormalizes the nonrelativistic
level spacing at $\mathcal{O}(m^{2}/\Ep^{2})$. The light-like deformation
combines both mechanisms. These results provide a compact analytic
benchmark for testing how different DSR realizations imprint themselves
on relativistic bound-state spectra and their nonrelativistic limits.
Future work may extend the present analysis to higher dimensions,
include external fields, or explore thermodynamic and information-theoretic
indicators within the same projective DSR framework.

\end{document}